\documentclass[doublecol]{epl2}

\title{Vector magnetic field sensing by single nitrogen vacancy center in
diamond}
\author{X.-D. Chen\inst{1} \and F.-W. Sun\inst{1}\thanks{E-mail: \email{fwsun@ustc.edu.cn}} \and C.-L. Zou\inst{1} \and J.-M. Cui\inst{1} \and L.-M. Zhou\inst{1} \and G.-C. Guo\inst{1}}

\institute{
\inst{1}Key Lab of Quantum Information, University of Science and Technology of
China - Hefei 230026, P.R.China}

 {Color centers and other defects}
	{Nuclear states and interactions}
	{Optically detected magnetic resonance (ODMR)}

\abstract{
In this Letter, we proposed and experimentally demonstrated a method to detect vector magnetic field with a single nitrogen vacancy (NV) center in diamond. The magnetic field in parallel with the axis of the NV center can be obtained by detecting the electron Zeeman shift, while the Larmor precession of an ancillary nuclear spin close to the NV center can be used to measure the field perpendicular to the axis. Experimentally, both the Zeeman shift and Larmor precession can be measured through the fluorescence from the NV center. By applying additional calibrated magnetic fields, complete information of the vector magnetic field can be achieved with such a method. This vector magnetic field detection method is insensitive to temperature fluctuation and it can be applied to nanoscale magnetic measurement.
}

\begin{document}

\maketitle

\section{Introduction}
The magnetic field plays a significant role in various regions such as material science and quantum physics since it can affect the motions of charged particles and the dynamics of electrons and nuclei spin states. Therefore, to precisely detect the magnetic field is becoming more and more important. Various magnetic detection techniques have been developed based on different systems including scanning Hall probe microscope \cite{apl-hallprobe}, magnetic resonance force microscope \cite{magneticRevModPhys.67.249}, optical magnetometry \cite{optmag} and scanning superconducting quantum interference device (SQUID) \cite{squid}. Very recently, nitrogen vacancy (NV) centers in diamond were proposed to detect magnetic field with nanoscale spatial resolution \cite{jel-nature2008-1,lukin-natphys2008-1,lukin-nature2008-1}. The general idea of NV magnetometer is that the electron spin energy levels change with external magnetic field due to electron Zeeman effect. The great advantages of such a magnetometer are that the electron spin states in NV center possess long coherence time at room temperature and they can be optically initialized and detected \cite{elestr-prb2006}. Therefore, the NV magnetometry is very potential for future practical applications. It is not only convenient to be operated without cryostat as required in SQUID, but also very sensitive to weak magnetic field, even the magnetic field produced by single remote nuclear spin\cite{Wrachtrup-sen-nucl,prl-sen-nucl-lukin,prl-sen-nucl-hanson}.

It is also important and necessary to detect the vector magnetic field with a single NV center. The interest in the detection of vector magnetic field has been driven by numerous applications. For example, it can provide both translation and rotation information of nanoparticle in biological cell \cite{cellnv,Hollenberg-nn-2011}. However, the application of NV center in magnetic measurement is restricted by the fact that the electron Zeeman shift mainly depends on the external magnetic field component along the defect's symmetry axis, especially for weak magnetic field. Though several NV centers with different orientations have been used to detect the local magnetic field vector \cite{Awschalom-apl2010-magnetic}, they were separated by hundreds of nanometers to be optically distinguished, which will reduce the spatial resolution.

Here, we proposed a method to detect the vector magnetic field by a single NV center, with the assistance of $^{15}$N nuclear spin. Since the distance between nuclear spin and electron spin is less than $1$ nm, high spatial resolution can be achieved. The magnetic field in parallel with the axis of $^{15}$NV center is obtained by detecting the electron Zeeman shift. To get the magnetic field perpendicular to the axis, the spin-selective pulse is used to measure nuclear spin state, which undergoes Larmor precession in the magnetic field. The results of electron spin Zeeman effect and nuclear spin Larmor precession can be detected through the fluorescence from the $^{15}$NV center. Therefore, the value of magnetic field is determined, so is the angle between NV center symmetry axis and the magnetic field. Complete vector magnetic field can be determined by introducing calibrated magnetic fields. Such a vector magnetic field sensor can be used for future studies of nano-material and biological system.

\begin{figure}[tbp]
\centering
\includegraphics[width=7cm]{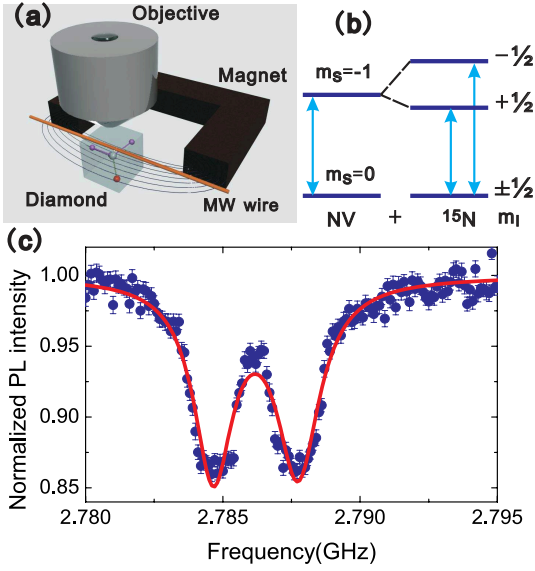}
\caption{(Color online) (a) The schematic of experimental setup of single NV center magnetometry in diamond. (b) Level structure of $^{15}$NV center. (c) The ODMR spectrum of $^{15}$NV center with an unknown magnetic field. The spectrum reveals the hyperfine structure of $^{15}$N.}
\label{ODMR}
\end{figure}

\section{Model}
The NV center in diamond consists of a substitutional
nitrogen atom and an adjacent vacancy with $\mathrm{{C_{3v}}}$ symmetry. The ground state of NV electron spin is a triplet with $S=1$. The Hamiltonian of NV center electron spin reads
\begin{equation}
H_{e}=DS_{z}^{2}+g_{e}\beta _{e}\mathbf{B}\cdot \mathbf{S}.  \label{eq1}
\end{equation}
Here, $\emph{z}$ denotes the NV center symmetry axis. $\mathbf{S}$ is the electron spin vector operator. $g_{e}$ and $\beta _{e}$ are the electron $g$ factor and Bohr magneton, respectively. $D\approx 2.87 GHz$ is zero field split (ZFS) between sublevels $m_{s}=0$ and $m_{s}=\pm 1$  at room temperature (Fig.\ref{ODMR}(b)). These spin states can be manipulated by microwave and detected through fluorescence intensity. Due to Zeeman effect, an external static magnetic field will cause the shifts of electron spin states. Usually, without the knowledge of the direction of external magnetic field, we can not distinguish the electron spin transitions between $\Delta m_{s} = + 1$ and $\Delta m_{s}= - 1$ only from the resonant spectrum. But we can definitely get two different resonant frequencies $\omega_{\pm}$ with $\omega_{+}>\omega{-}$. Under weak field condition with $|\mathbf{B}|\ll D/g_{e}\beta _{e}$, the shifts caused by magnetic can be expressed  as
\begin{equation}  \label{zeeman2}
\Delta\omega_{\pm} \approx \pm g_{e}\beta _{e}|B_{z}| + \frac{1}{2}%
g_{e}^{2}\beta _{e}^{2}B_{\perp }^{2} \frac{3 D \mp g_{e}\beta _{e}|B_{z}|}{D^{2}- g_{e}^{2} \beta _{e}^{2} |B_{z}|^{2}},
\end{equation}
where $B_{\perp}= \sqrt{B_{x}^{2}+B_{y}^{2}}$. Obviously, the second term is much smaller than the first term. So the Zeeman shift is mainly determined by $B_{z}$.

However, the electron spin of NV center is not isolated in diamond. Usually three species of
nuclear spins are found to be coupled with the electron spins of NV
center: $^{14}$N or $^{15}$N atom associated with the NV center, and randomly placed $^{13}$%
C atoms. When the distance between nuclear spin and electron spin of NV center is small, the nuclear spin will
affect the dynamics of electron spin significantly.
The Hamiltonian of a nearby nuclear spin and its hyperfine interaction with the electron
spin are
\begin{eqnarray}
H_{n} &=&-g_{n}\beta _{n}\mathbf{B}\cdot \mathbf{I}+\mathbf{I}\cdot \mathbf{P%
}\cdot \mathbf{I},  \label{eq1-2} \\
H_{i} &=&\mathbf{S}\cdot \mathbf{A}\cdot \mathbf{I}.
\end{eqnarray}%
Here, $g_{n}$ and $\beta _{n}$ are the nuclear $g$ factor and magneton respectively. $%
\mathbf{A}$ is the hyperfine interaction tensor. And $\mathbf{P}$ is the
nuclear quadrupole interaction tensor for $\emph{I}\geq 1$.

Our magnetic measurement idea is that $\mathbf{B}$ can affect the nuclear spin indirectly through the hyperfine interaction, then the dynamics of nuclear spin is more magnetic sensitive than the nuclear Zeeman effect. Here, a single $^{15}$NV was chosen to present the measurement method. The $%
^{15}$NV center can be found in natural diamond. It also can be created
artificially by ion implantation \cite{N14-APL2006}, which makes it feasible
for future application. In such a composite system, the Larmor precession of $
^{15}$N is affected not only by the external magnetic field, but also the electron spin state\cite%
{lukin-sci2006dynamics,lukin-sci2007-register}. Here, we focus on the nuclear free precession with $m_{s}=0$. The Larmor frequency is expressed as
\begin{equation}  \label{l}
\omega_{L}=|g_{n}\beta _{n}\mathbf{B}_{eff}|
\end{equation}
with $\mathbf{B}_{eff}=\mathbf{B}\cdot \mathbf{g}_{0}$. The $\mathbf{g}%
_{0}$ tensor can be calculated by second order perturbation theory \cite%
{lukin-farfield,lukin-sci2006dynamics}:
\begin{equation}
\mathbf{g}_{0}\approx \left(
\begin{array}{ccc}
1 & 0 & 0 \\
0 & 1 & 0 \\
0 & 0 & 1%
\end{array}%
\right) +\frac{2g_{e}\beta _{e}}{g_{n}\beta _{n}D}\left(
\begin{array}{ccc}
A_{xx} & A_{xy} & A_{xz} \\
A_{yx} & A_{yy} & A_{yz} \\
0 & 0 & 0%
\end{array}%
\right) ,  \label{eq2}
\end{equation}%
where $A_{i,j}$ is the element of hyperfine interaction tensor $\mathbf{A}$.
As the nuclear spin $^{15}$N lies on NV center's symmetry axis, the
interaction tensor has diagonal form\cite{lukin-sci2007-register} $%
A_{ij}\equiv A_{ij}\delta _{i,j}$. Therefore the relation between nuclear
spin Larmor frequency and external magnetic field can be simplified as
\begin{equation}  \label{lam}
\omega_{L} \approx | g_{n} \beta _{n} | \sqrt{\alpha
_{x}^{2}B_{x}^{2}+\alpha_{y}^{2}B_{y}^{2}+B_{z}^{2}},
\end{equation}
with the coefficient $\alpha _{i}=1+\frac{2g_{e}\beta _{e}}{g_{n}\beta
_{n}D}\times A_{ii}$.

These measurable frequencies $\Delta\omega_{\pm}$ and $\omega_{L}$ are
functions of $|B_{z}|$ and $|B_{\perp}|$. Therefore we can solve the
absolute values of total magnetic field and polar angle between the vector
and the NV axis. The nuclear Zeeman effect is ignored in $%
\Delta\omega_{\pm}$ since it is much smaller than the electron Zeeman effect as $%
\beta_{e}/\beta _{n}\approx 1836$.

\begin{figure}[tbp]
\centering
\includegraphics[width=7cm]{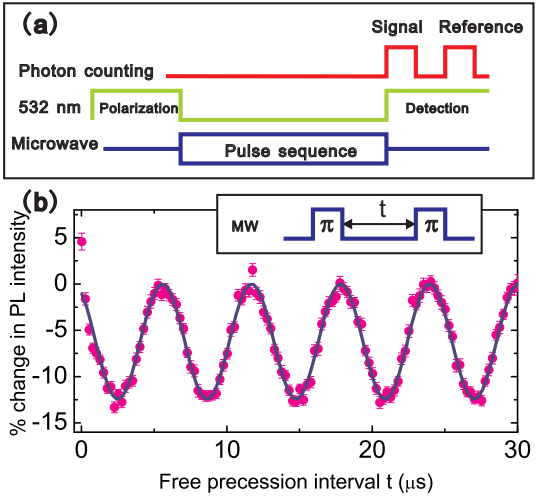}
\caption{(Color online) (a) Pulse sequence in the experiment. (b) $^{15}$%
N nuclear spin precession. Pink dots are the experimental results, and the
blue solid line is cosine function fitting. }
\label{bper}
\end{figure}

\section{Experiment and results}
The experiment was performed on a (100) CVD diamond plate
from Element Six, with nitrogen concentration lower than 5 ppb. The experimental setup is shown in Fig.\ref%
{ODMR}(a). All measurements were carried out at room temperature with a home
built optical confocal scanning system. Microwave was transmitted through a 15 $\mu $m
diameter copper wire, and a permanent magnet was placed near the sample to
provide external static magnetic field.

As shown in Fig.\ref{ODMR}(c), the transition frequencies between spin states can be
measured through optically detected magnetic resonance (ODMR) \cite%
{ODMR-science1997,jel-nature2008-1}. When the applied microwave was resonant
with one of the transition frequencies, the fluorescence intensity
would show a decrease. More precise results can be obtained by other methods
like Ramsey fringes \cite{Dutt-natnano2011}. The $^{15}$N atom with nuclear
spin $\emph{I}=1/2$ split the electron spin states, as shown
in Fig.\ref{ODMR}(b). Therefore, we can observe two valleys in Fig.\ref{ODMR}(c)
for one electron spin transition branch ($m_{s}=0\rightarrow
m_{s}=-1$ or $m_{s}=0\rightarrow m_{s}=+1$). The Lorentzian fitting revealed two
resonant frequencies 2.78465(2) GHz and 2.78774(2) GHz, corresponding to
different nuclear spin states. Comparing with the zero magnetic field resonant
frequencies, we got $\Delta\omega_{-}=- 85.26(6) MHz$ under the external static magnetic field.

The free precession signal of $^{15}$N nuclear spin was obtained by two nuclear spin selective microwave $\pi $ pulses. The pulse sequences and results are shown in Fig.\ref{bper}(a)
and (b). The electron spin of NV center was first polarized to $m_{s} = 0$ by $3 \mu s$ green laser
pulses. Without nuclear spin polarization, the
nuclear spin was assumed to be at thermal equilibrium state with the state density
matrix $P_{\uparrow }|0,+\frac{1}{2}\rangle \langle 0,+\frac{1}{2}%
|+P_{\downarrow }|0,-\frac{1}{2}\rangle \langle 0,-\frac{1}{2}| $ following
the basis $|m_{s},m_{I}\rangle \langle m_{s},m_{I}|$. After the first
selective $\pi $ pulse (e.g. on the transition $|0,+\frac{1}{2}\rangle
\rightarrow |-1,+\frac{1}{2}\rangle $), the density matrix became $%
P_{\uparrow }|-1,+\frac{1}{2}\rangle \langle -1,+\frac{1}{2}|+P_{\downarrow
}|0,-\frac{1}{2}\rangle \langle 0,-\frac{1}{2}|$. During the free precession
time $t$, the nuclear spin with $m_{s}=0$ periodically changed
the direction, while the split between nuclear spin states in the $m_{s}=-1$ subspace
prevented the precession of nuclear spin\cite%
{ODMR13Cspectr-prb,lukin-sci2007-register}. The second microwave $\pi $ pulse mapped
the information of nuclear spin onto the electron spin. At last, another
green laser pulse was used to read out the electron spin. The detected
photoluminescence intensity oscillated as $ I(t)\propto P_{\downarrow
}cos( 2 \pi \omega _{L}t)$ with Larmor frequency $\omega _{L}$.
Fitting the precession signal by cosine function, we got $\omega _{L}=
$163.2(3) KHz in Fig. \ref{bper}(b).

\begin{figure}[tbp]
\centering
\includegraphics[width=8.5cm]{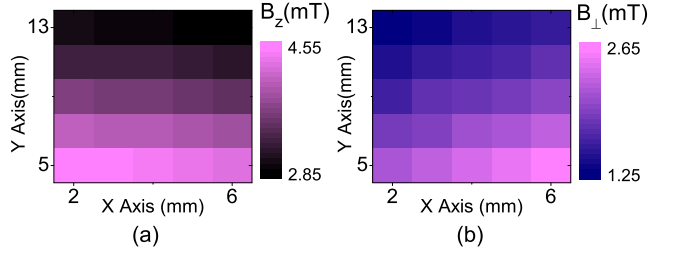}
\caption{(Color online) The magnetic field intensity imaging detected by a
single $^{15}$NV. (a) and (b) are the absolute values of $B_{z}$ and $B_{\perp}$, respectively, obtained through the electron Zeeman effect and $^{15}$N Larmor precession.}
\label{map}
\end{figure}

For $^{15}$N, the nuclear $g$ factor is about $-0.566$. The hyperfine
interaction parameters are $A_{xx}=A_{yy}=3.65(3)$ \textrm{MHz} and $%
A_{zz}=3.03(3)$ \textrm{MHz}\cite{hyperfineparameter-PhysRevB2009}, thus $\alpha _{x}=\alpha _{y}=-15.5(1)$. Substituting these parameters into
Eqs. (\ref{zeeman2}) and (\ref{lam}), we solved the absolute magnitude of the magnetic field as $%
|B_{z}| =3.129 mT$ and $|B_{\perp} | = 2.426 mT$. The second term in Eq. (2) was about $2.4 MHz$. It was small
but can not be neglected for high accuracy magnetometry. Utilizing this method, we presented the spatial distribution of vector magnetic field by moving the permanent magnet along X or Y axis (the laboratory coordinate frame). The results of $B_{z}$ and $B_{\perp }$ are shown in Fig. \ref{map}. As the angle between magnetic field and NV axis varies with position, the change of $B_{z}$ might be contrary to $B_{\perp }$ at the some locations.

\begin{figure}[tbp]
\includegraphics[width=7cm]{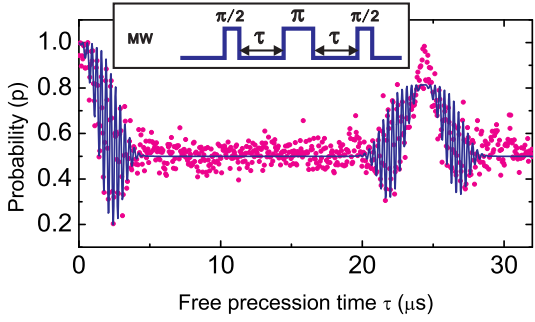}
\caption{(Color online) Experimental results of electron spin echo signal (pink points) and fitting with theoretical formula (blue line). Inset: the experimental control pulse for electron spin echo signal.}
\label{spinecho}
\end{figure}

For comparison, the electron spin echo was also measured to estimate $|\mathbf{B}|$. As shown in Fig. 4, the spin echo signal is modulated by both $^{15}$N spin and $^{13}$C spin bath. The coupling between $^{13}$C nuclear spin ensembles and NV electron spin contributes to the collapse and revival of the electron spin echo\cite{lukin-sci2006dynamics,spinechomodu-PR}, while the coupling between $^{15}$N and $^{15}$NV electron spin
contributes to the rapid oscillations. The result was fitted by the model including the hyperfine interaction with $^{15}$N as \cite{lukin-sci2006dynamics} $S=\frac{1}{2}+\frac{1}{2}(e^{-(\tau / \tau _{c})^{4}}+b\times e^{-((\tau -\tau _{re})/ \tau _{c})^{4}})\times \lbrack 1-a\times sin(2 \pi \omega _{1}\protect\tau /2)^{2}sin(2 \pi\omega _{2} \tau /2)^{2}]$, where $\omega_{1}$ and $\omega_{2}$ were Larmor frequencies of $^{15}$N nuclear spin with different electron spin states. The first revival peak was at $ \tau _{re} = 24.27(6) \mu s$, corresponding to the revival frequency $\omega_{revival}\approx 41.2 KHz$, which equaled to Larmor frequency of $^{13}$C. It corresponded to a magnitude of the field \textbf{$|\mathbf{B}|=\omega_{revival}/g_{^{13}C}\beta_{^{13}C} \approx 3.85 mT$}. This
agreed with the previous results, where $|\mathbf{B}|= \sqrt{%
B_{z}^{2} + B_{\perp}^{2}} \approx 3.96 mT$. Combining with electron Zeeman effect, the electron spin echo can also provide the information of vector magnetic field. However, the detection of spin echo is usually slower than that of nuclear free precession. Furthermore, $^{13}$C nuclear spins are randomly distributed around the NV center, and the nuclear spins close to the NV center will lead to very rapid oscillations in the spin echo signal. All of these make the revival peaks difficult to be located precisely and lead to detection error.

\section{Discussion}

Now we will discuss the sensitivity in Figs.\ref{ODMR} and \ref{bper}. The ODMR spectrum has Lorentzian shape, and there are two points on the spectrum with the highest sensitivity. The noise of resonant frequencies at these two points are $60 KHz$. Therefore the noise of $B_{z}$ caused by ODMR method is about $2 \mu T$.

To analyze the sensitivity of $B_{\perp}$, Eq.(\ref{lam}) is rewritten as
\begin{equation}\label{lam2}
 B_{\perp} \approx \frac{1}{\alpha_{x}} \sqrt{\frac{\omega_{L}^2}{|g_{n} \beta_{n}|^2}-B_{z}^{2}}.
\end{equation}
In this experiment the noise caused by nuclear spin precession is much larger, so only the error of $\omega_{L}$ is taken into account here. And the precession signal is well fitted by
\begin{equation}\label{signal}
    I(t)=I_{0}+I_{c} cos( 2 \pi \omega_{L} t)e^{-t/T_{0}},
\end{equation}
where $T_{0}=156 \mu s$ is the nuclear spin coherence time in the present measurement. The maximal sensitivity is obtained at $cos( 2 \pi \omega_{L} t) = 0$\cite{lukin-nature2008-1}. The uncertainty $\delta B_{\perp}$ is
\begin{equation}\label{uncer}
    \delta B_{\perp} = \delta I \frac{\partial B_{\perp}}{\partial I} \approx \delta I \frac{1}{2 \pi \alpha_{x} |g_{n} \beta_{n}| I_{c} t}e^{t/T_{0}}.
\end{equation}
After polarizing the nuclear spin by selective $\pi$ microwave pulse and optical excitation pulse (results not shown)\cite{lukin-sci2007-register}, the noise ratio is reduced to $\frac{\delta I}{I_{c}}\approx 0.045$. Then the error of $B_{\perp}$ with our method is about $1.8 \mu T$ at $t = T_{0}$. Therefore, the minimal detectable amplitude of magnetic field is at scale of $\mu T$ with our method, and the error of angle is lower than $0.05^{\circ}$. The NV center used in our experiment is very deep inside the crystal (about decades of micrometers), then only $3\times 10^{4}$ photons are detected per second. Using the implanted $^{15}$NV centers which are near the sample surface, the collection efficiency can be improved. And longer coherence time of $^{15}$N nuclear spin will also improve the sensitivity.

With the method in Ref.{\cite{jel-nature2008-1}}, where all electron spin
resonant frequencies' shifts $\Delta \omega _{\pm }$ are measured with ODMR, $|B_{z}|$ and $|B_{\perp }|$ can also be calculated by $\Delta\omega_{+}-\Delta\omega _{-}$ and $\Delta\omega _{+}+\Delta\omega _{-}$. In this way, the error $\delta B_{\perp} \approx D \frac{\delta \omega_{+}+\delta \omega_{-}}{6 g_{e}^{2} \beta_{e}^{2} B_{\perp}}$,  will be about $30 \mu T$ with $B_{\perp}= 2.4 mT$. It is higher than our method. Further more, the ODMR method is affected by the fluctuation of temperature. If the temperature changes $1$ \textrm{K} at room
temperature, D will shift about $84$ \textrm{KHz }\cite{temshift}. Then $%
\omega _{+}+\omega _{-}$ changes $168$ \textrm{KHz}, and the corresponding error
of $B_{\perp }$ could be more than $100 \mu T$ in weak magnetic situation. In
contrast, the Larmor frequency of $^{15}$N is mainly determined by $B_{\perp
}$, and it is insensitive to temperature fluctuation.

\begin{figure}[tbp]
\includegraphics[width=7cm]{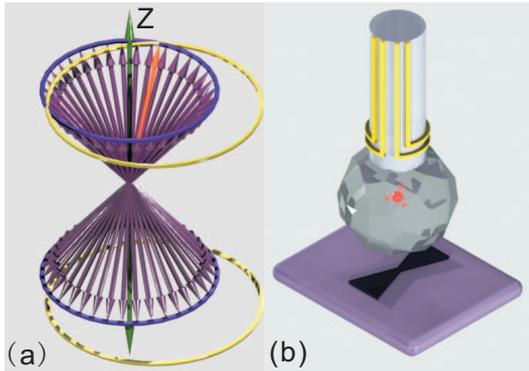}\newline
\caption{ (Color online) (a) The results without calibrated field defines two rings (blue) in the vector space, as the azimuthal angle is not sure. Another two rings (yellow) can be obtained by applying a calibrated field. The intersections of the results with and without calibrated field indicate that the real direction (orange) is in the two possible directions. (b) A proposal schematic of a compact magnetic probe tip.}
\label{vector}
\end{figure}

Although only the magnitude and polar angle of the vector magnetic field are obtained so far, it is possible to deduce all the vector information by additional calibrated fields. As shown in Fig.\ref{vector} (a), all possible magnetic vectors started at a fixed point in the vector space with the endpoints limited on two rings (in blue). By adding the additional calibrated field, we can get another two rings of the possible endpoints of the detected magnetic. The two yellow rings in Fig.\ref{vector} (a) show the measured results by subtracting the additional calibrated field. If the calibrated field is neither perpendicular nor parallel to the NV axis, there would be only two possible directions of the unknown magnetic field left. In this way, two different calibrated fields are needed to calculate the complete magnetic field vector.

We also propose a near field scanning vector magnetometry with nanoscale spatial resolution, as shown in Fig. 5(b). A single $^{15}$NV center is fixed at the tip of atomic force microscope (AFM) probe. There are metal strip lines on the tip which can generate calibrated magnetic field with fixed direction by applying specific current. As the distance between nuclear spin and electron spin is lower than 1 nm, the spatial resolution should be several nanometers, which is limited by the AFM. Using this scanning magnetometry, the three-dimensional vector magnetic field distribution can be measured with high precision and spatial resolution, which can be used to study the dynamics of biological systems, and probe the properties of nanostructure which consists of magnetic materials.

\section{Conclusion}
In summary, we proposed and demonstrated a vector magnetic field magnetometry by a single NV center with an ancillary nuclear spin. In this composite spin system, the frequency shift of electron spin and the Larmor frequency of nuclear spin can be detected optically when the NV center is placed in external static magnetic field. Then, the amplitudes of magnetic field parallel and perpendicular to the axis of NV are determined. Further more, by applying calibrated magnetic fields with defined directions, complete information of the vector can be measured. In future work, the precision of measurement can be improved by methods like dynamic decoupling. With high spatial resolution system like AFM, the magnetic properties of nanoscale particle can be detected.

\acknowledgments
This work was supported by the National Fundamental Research Program of
China under Grant No. 2011CB921200, the Knowledge Innovation Project of
Chinese Academy of Sciences (60921091), National Natural Science Foundation
of China under Grant No.11004184.



\begin{thebibliography}{10}
\expandafter\ifx\csname url\endcsname\relax\def\url#1{\texttt{#1}}\fi

\bibitem{apl-hallprobe}
\Name{Chang A.~M., Hallen H.~D., Harriott L., Hess H.~F., Kao H.~L., Miller
  R.~E., Wolfe R., van~der Ziel J. \and Chang T.~Y.} \REVIEW{Appl. Phys.
  Lett.}{61}{1992}{1974}.

\bibitem{magneticRevModPhys.67.249}
\Name{Sidles J.~A., Garbini J.~L., Bruland K.~J., Rugar D., Z\"uger O., Hoen S.
  \and Yannoni C.~S.} \REVIEW{Rev. Mod. Phys.}{67}{1995}{249}.

\bibitem{optmag}
\Name{Budker D. \and Romalis M.} \REVIEW{Nature Phys.}{3}{2007}{227}.

\bibitem{squid}
\Name{Bending S.~J.} \REVIEW{Adv. Phys.}{48}{1999}{449}.

\bibitem{jel-nature2008-1}
\Name{Balasubramanian G., Chan I.~Y., Kolesov R., Al-Hmoud M., Tisler J., Shin
  C., Kim C., Wojcik A., R.Hemmer P., Krueger A., Hanke T., Leitenstorfer A.,
  Bratschitsch R., Jelezko F. \and Wrachtrup J.}
  \REVIEW{Nature}{455}{2008}{648}.

\bibitem{lukin-natphys2008-1}
\Name{Taylor J.~M., Cappellaro P., Childress L., Jiang L., Budker D., Hemmer
  P.~R., Yacoby A., Walsworth R. \and Lukin M.~D.} \REVIEW{Nature
  Phys.}{4}{2008}{810}.

\bibitem{lukin-nature2008-1}
\Name{Maze J.~R., Stanwix P.~L., Hodges J.~S., Hong S., Taylor J.~M.,
  P.Capperllaro, L.Jiang, Dutt M., Togan E., Zibrov A.~S., Yacoby A., Walsworth
  R.~L. \and Lukin M.~D.} \REVIEW{Nature}{455}{2008}{644}.

\bibitem{elestr-prb2006}
\Name{Manson N.~B., Harrison J.~P. \and Sellars M.~J.} \REVIEW{Phys. Rev.
  B}{74}{2006}{104303}.

\bibitem{Wrachtrup-sen-nucl}
\Name{Zhao N., Honert J., Schmid B., Klas M., Isoya J., Markham M., Twitchen
  D., Jelezko F., Liu R.-B., Fedder H. \and Wrachtrup J.} \REVIEW{Nature
  Nanotechnology}{7}{2012}{657}.

\bibitem{prl-sen-nucl-lukin}
\Name{Kolkowitz S., Unterreithmeier Q.~P., Bennett S.~D. \and Lukin M.~D.}
  \REVIEW{Phys. Rev. Lett.}{109}{2012}{137601}.

\bibitem{prl-sen-nucl-hanson}
\Name{Taminiau T.~H., Wagenaar J. J.~T., van~der Sar T., Jelezko F.,
  Dobrovitski V.~V. \and Hanson R.} \REVIEW{Phys. Rev.
  Lett.}{109}{2012}{137602}.

\bibitem{cellnv}
\Name{Hall L.~T., Beart G. C.~G., Thomas E.~A., Simpson D.~A., McGuinness
  L.~P., Cole J.~H., J.~H.~Manton R. E.~S., Jelezko F., Wrachtrup J., Petrou S.
  \and Hollenberg L. C.~L.} \REVIEW{Sci. Rep.}{2}{2012}{401}.

\bibitem{Hollenberg-nn-2011}
\Name{McGuinness L.~P., Yan Y., Stacey A., Simpson D.~A., Hall L.~T., Maclaurin
  D., Prawer S., Mulvaney P., Wrachtrup J., Caruso F., Scholten R.~E. \and
  Hollenberg L. C.~L.} \REVIEW{Nat. Nano.}{6}{2011}{358}.

\bibitem{Awschalom-apl2010-magnetic}
\Name{Maertz B.~J., Wijnheijmer A.~P., Fuchs G.~D., Nowakowski M.~E. \and
  Awchalom D.~D.} \REVIEW{Appl. Phys. Lett.}{96}{2010}{092504}.

\bibitem{N14-APL2006}
\Name{Rabeau J., Reichart P., Tamanyan G., Jamieson D., Prawer S., Jelezko F.,
  Gaebel T., Popa I., Domhan M. \and Wrachtrup J.} \REVIEW{Appl. Phys.
  Lett.}{88}{2006}{023113}.

\bibitem{lukin-sci2006dynamics}
\Name{Childress L., Gurudev~Dutt M.~V., Taylor J.~M., Zibrov A.~S., Jelezko F.,
  Wrachtrup J., Hemmer P.~R. \and Lukin M.~D.}
  \REVIEW{Science}{314}{2006}{281}.

\bibitem{lukin-sci2007-register}
\Name{Dutt M. V.~G., Childress L., Jiang L., Togan E., Maze J., Jelezko F.,
  Zibrov A.~S., Hemmer P.~R. \and Lukin M.~D.}
  \REVIEW{Science}{316}{2007}{1312}.

\bibitem{lukin-farfield}
\Name{Maurer P.~C., Maze J.~R., Stanwix P.~L., Jiang L., Gorshkov A.~V., Zibrov
  A.~A., Harke B., Hodges J.~S., Zibrov A.~S., Yacoby A., Twitchen D., Hell
  S.~W., Walsworth R.~L. \and Lukin M.~D.} \REVIEW{Nature Phys.}{6}{2010}{912}.

\bibitem{ODMR-science1997}
\Name{Gruber A., Dr$\ddot{a}$benstedt A., Tietz C., Fleury L., Wrachtrup J.
  \and C.von B.} \REVIEW{Science}{276}{1997}{2012}.

\bibitem{Dutt-natnano2011}
\Name{Nusran N.~M., Momeen M.~U. \and Dutt M. V.~G.} \REVIEW{Nature
  Nanotech.}{7}{2012}{109}.

\bibitem{ODMR13Cspectr-prb}
\Name{Dr\'eau A., Maze J.-R., Lesik M., Roch J.-F. \and Jacques V.}
  \REVIEW{Phys. Rev. B}{85}{2012}{134107}.

\bibitem{hyperfineparameter-PhysRevB2009}
\Name{Felton S., Edmonds A.~M., Newton M.~E., Martineau P.~M., Fisher D.,
  Twitchen D.~J. \and Baker J.~M.} \REVIEW{Phys. Rev. B}{79}{2009}{075203}.

\bibitem{spinechomodu-PR}
\Name{Rowan L.~G., Hahn E.~L. \and Mims W.~B.} \REVIEW{Phys.
  Rev.}{137}{1965}{A61}.

\bibitem{temshift}
\Name{Chen X.-D., Dong C.-H., Sun F.-W., Zou C.-L., Cui J.-M., Han Z.-F. \and
  Guo G.-C.} \REVIEW{Appl. Phys. Lett.}{99}{2011}{161903}.

\end{thebibliography}
\end{document}